\documentclass[draftclsnofoot, onecolumn]{IEEEtran}

\usepackage[utf8]{inputenc} 
\usepackage[T1]{fontenc}
\usepackage{microtype}
\usepackage{url}
\usepackage{ifthen}
\usepackage[noadjust]{cite}
\usepackage[cmex10]{amsmath}
\usepackage{mathtools}
\usepackage{dsfont}
\usepackage{xcolor}
\usepackage{mathdots}
\usepackage{amsfonts}
\usepackage{cite}
\usepackage[capitalise]{cleveref}
\usepackage{hhline}

\DeclareMathOperator{\RS}{RS}
\DeclareMathOperator{\GRS}{GRS}

\DeclareMathOperator{\spn}{span}
\DeclareMathOperator{\wt}{wt}
\DeclareMathOperator{\supp}{supp}

\renewcommand{\sec}{\mathrm{sec}}
\newcommand{\F}{\mathbb{F}}
\renewcommand{\tilde}{\widetilde}

\newtheorem{theorem}{Theorem}
\newtheorem{conjecture}{Conjecture}
\newtheorem{lemma}{Lemma}
\newtheorem{proposition}{Proposition}
\newtheorem{example}{Example}

\begin{document}
\title{Flexible Field Sizes in Secure Distributed Matrix Multiplication via Efficient Interference Cancellation}

\author{%
    \IEEEauthorblockN{Okko~Makkonen}\\
    \IEEEauthorblockA{Department of Mathematics and Systems Analysis\\
    Aalto University\\
    Espoo, Finland\\
    Email: okko.makkonen@aalto.fi}
}

\maketitle

\begin{abstract}
In this paper, we propose a new secure distributed matrix multiplication (SDMM) scheme using the inner product partitioning.
We construct a scheme with a minimal number of workers and no redundancy, and another scheme with redundancy against stragglers.
Unlike previous constructions in the literature, we do not utilize algebraic methods such as locally repairable codes or algebraic geometry codes.
Our construction, which is based on generalized Reed--Solomon codes, improves the flexibility of the field size as it does not assume any divisibility constraints among the different parameters.
We achieve a minimal number of workers by efficiently canceling all interference terms with a suitable orthogonal decoding vector.
Finally, we discuss how the MDS conjecture impacts the smallest achievable field size for SDMM schemes and show that our construction almost achieves the bound given by the conjecture.
\end{abstract}

\section{Introduction}
\label{sec:introduction}

Secure distributed matrix multiplication (SDMM) is a way of distributing the computation of a matrix product to a distributed system of worker nodes, while providing resiliency against slow and unresponsive workers (stragglers) and data security against colluding workers.
SDMM was first introduced by Chang and Tandon in \cite{chang2018capacity} and has seen a number of papers improving on their construction \cite{yang2018secure,kakar2019capacity,aliasgari2020private,d2020gasp,d2021degree,jia2021cross}.
SDMM schemes utilize many techniques from coding theory and secret sharing, which means that they naturally operate over finite fields.

The way to distribute the computation is to break the problem down to smaller problems.
This is done by partitioning the two matrices to smaller pieces such that the entire product can be assembled from the smaller products.
In this paper we focus on the so-called inner product partitioning where the product can be computed as an inner product of the partition vectors.

If all responses from the worker nodes are needed, then just one node being slow or unresponsive will slow down the whole computation.
This problem, known as the \emph{straggler problem}, can be reduced by making the system sufficiently redundant such that some of the workers' responses may be ignored.
This redundancy is provided through methods in coding theory, where these straggling workers may be seen as erasures in the codewords.
Coding theory can also be used to protect against errors in the responses coming from so-called \emph{Byzantine workers}, which have been addressed \cite{makkonen2022general,byrne2023straggler}.

In case the matrices contain sensitive information, it is important for the distribution process not to leak any secret information to the worker nodes.
This is achieved through secret sharing, where coded randomness is inserted to the computation such that any sufficiently few nodes will not be able to decode the original matrices.
In particular, information-theoretic security is required for secure distributed matrix multiplication.

Many constructions in the literature use techniques from algebraic coding theory, such as locally repairable codes in \cite{lopez2022secure}, algebraic geometry codes over Hermitian curves and Kummer extensions in \cite{machado2023hera} and \cite{makkonen2023algebraic}, and so-called discrete Fourier transform (DFT) codes in \cite{mital2022secure}.
These constructions assume certain properties of the finite fields, such as the existence of primitive $N$th roots of unity in \cite{mital2022secure}, or large field extensions in \cite{machado2021field}.
Furthermore, the authors of \cite{machado2023hera} are interested in finding SDMM schemes over small fields by using codes over high genus algebraic curves.
In this paper, we construct SDMM schemes without relying on assumptions on the field, other than that the field has to be sufficiently large ($q \geq N$, where $N$ is the number of workers) for the existence of certain MDS codes.
Our construction is enabled by efficient interference cancellation, where all noise terms are removed with a single linear combination, as well as the observation that computing any non-degenerate bilinear form is sufficient for the inner product partitioning.
This method does not increase the encoding or decoding complexity.
We also discuss how the field size has to be roughly $N$ due to the MDS conjecture.

This paper is organized as follows.
In \cref{sec:preliminaries}, we give some preliminaries on linear codes and generalized Reed--Solomon codes.
In \cref{sec:SDMM}, we introduce the system model for secure distributed matrix multiplication and give some examples.
In \cref{sec:construction}, we provide our construction utilizing Reed--Solomon codes with flexibility in the chosen finite field.
Finally, in \cref{sec:comparison}, we compare our construction against some competing schemes in the literature.

\section{Preliminaries}
\label{sec:preliminaries}

Let $\F_q$ denote the finite field with $q$ elements, $\F_q^* = \F_q \setminus \{0\}$, and $[n] = \{1, \dots, n\}$.
We consider elements of $\F_q^n$ as row vectors.

\subsection{Linear Codes}
\label{ssec:linear_codes}

For a vector $x \in \F_q^n$ we define the \emph{support} and \emph{weight} by $\supp(x) = \{i \in [n] : x_i \neq 0\}$ and $\wt(x) = \lvert \kern 0.1em \supp(x) \rvert$.
A $k$\nobreakdash-dimensional subspace $\mathcal{C}$ of $\F_q^n$ is said to be an $[n, k]$ \emph{linear~code}.
If $G$ is a $k \times n$ matrix whose rows span the subspace $\mathcal{C}$, then $G$ is said the be a \emph{generator matrix} of $\mathcal{C}$.
The \emph{minimum distance} of a code $\mathcal{C}$ is
\begin{equation*}
    d_\mathrm{min} = \min \{ \wt(c) : c \in \mathcal{C} \setminus \{0\} \}.
\end{equation*}
The famous Singleton bound states that $d_\mathrm{min} \leq n - k + 1$ and codes that meet this bound with equality are said to be \emph{maximum distance separable (MDS)}.
Another characterization of MDS codes is that any $k \times k$ submatrix of the generator matrix $G$ is invertible.
Simple examples of MDS codes are the $[n, 1]$ repetition code and the $[n, n - 1]$ single parity check code.
The following proposition states that for $k \geq q + 1$, the longest possible MDS codes are single parity check codes.

\begin{proposition}[\cite{bush1952orthogonal}]
\label{prop:MDS_large_dimension}
Let $\mathcal{C}$ be a linear $[n, k]$ MDS code over $\F_q$ with $k \geq q + 1$.
Then, $n \leq k + 1$.
\end{proposition}

The following conjecture, which was stated by Segre in 1955 \cite{segre1955curve}, states that the longest possible (non-trivial) linear MDS codes have length roughly equal to the alphabet size.
Several special cases of the conjecture have been proven, see \cite{ball2012sets,ball2012sets2,chowdhury2015inclusion}.

\begin{conjecture}[MDS conjecture {\cite{segre1955curve}}]
\label{conj:MDS}
Let $\mathcal{C}$ be a linear $[n, k]$ MDS code over $\F_q$ with $2 \leq k \leq q$.
Then $n \leq q + 1$, except if $q$ is even and $k = 3$ or $k = q - 1$, in which case $n \leq q + 2$.
\end{conjecture}

The \emph{star product} of two length $n$ linear codes $\mathcal{C}$ and $\mathcal{D}$ is the linear code generated by the coordinatewise products of their codewords, \emph{i.e.},
\begin{equation*}
    \mathcal{C} \star \mathcal{D} = \spn\{ c \star d \mid c \in \mathcal{C}, d \in \mathcal{D} \},
\end{equation*}
where $(c \star d)_i = c_id_i$.

The \emph{dual} of a linear code $\mathcal{C}$ is a linear code $\mathcal{C}^\perp$ whose codewords are orthogonal to all the codewords in $\mathcal{C}$ with respect to the standard inner product, \emph{i.e.},
\begin{equation*}
    \mathcal{C}^\perp = \{x \in \F_q^n : x \cdot c = 0~\forall c \in \mathcal{C} \}.
\end{equation*}
If $\mathcal{C}$ is $k$\nobreakdash-dimensional, then $\mathcal{C}^\perp$ is $(n - k)$\nobreakdash-dimensional and $\mathcal{C} = (\mathcal{C}^\perp)^\perp$.
Furthermore, if $\mathcal{C}$ is MDS, then $\mathcal{C}^\perp$ is also MDS.

\subsection{Generalized Reed--Solomon Codes}
\label{ssec:GRS_codes}

A famous class of linear MDS codes can be constructed from evaluations of polynomials.
In particular, let $\alpha \in \F_q^n$ contain distinct entries, $\nu \in (\F_q^*)^n$, and let $\F_q[x]^{< k}$ denote the vector space of polynomials of degree $< k$.
The $k$\nobreakdash-dimensional \emph{generalized Reed--Solomon (GRS)} code is defined as
\begin{equation*}
    \GRS_k(\alpha, \nu) = \{(\nu_1 f(\alpha_1), \dots, \nu_n f(\alpha_n)) \mid f(x) \in \F_q[x]^{< k} \}.
\end{equation*}
It is well-known that GRS codes are MDS.
The elements of $\alpha$ and $\nu$ are known as evaluation points and column multipliers, respectively.
If $\mathds{1}$ is the all-ones vector, then we write $\RS_k(\alpha) = \GRS_k(\alpha, \mathds{1})$.
The length of a GRS code is $n \leq q$ due to the fact that the evaluation points have to be distinct points of $\F_q$.
Therefore, these codes do not quite achieve the bound of $n \leq q + 1$ (or $n \leq q + 2$) given by the MDS conjecture.

The star product of GRS codes defined on the same evaluation points can be computed with
\begin{equation*}
    \GRS_k(\alpha, \nu) \star \GRS_\ell(\alpha, \mu) = \GRS_{\min\{k + \ell - 1, n\}}(\alpha, \nu \star \mu).
\end{equation*}
The dual of an $(n - 1)$\nobreakdash-dimensional Reed--Solomon code is a one-dimensional MDS code, \emph{i.e.}, generated by some full-weight $\omega \in \F_q^n$.
The dual of a Reed--Solomon code can be computed from the following equation
\begin{equation*}
    (\RS_k(\alpha) \star \RS_{n-k}(\alpha))^\perp = \RS_{n-1}(\alpha)^\perp = \spn\{\omega\}.
\end{equation*}
Therefore,
\begin{equation*}
    0 = \sum_{i=1}^{N} \omega_i (c \star d)_i = \sum_{i=1}^{N} c_i (\omega_i d_i),
\end{equation*}
where $c \in \RS_k(\alpha)$ and $d \in \RS_{n - k}(\alpha)$.
Hence, $\RS_k(\alpha)^\perp = \GRS_{n-k}(\alpha, \omega)$.
Notice that $\omega$ only depends on the evaluation points $\alpha$ and not the dimension $k$.
In fact, the column multipliers may be written explicitly as $\omega_i = (\prod_{\substack{j=1, j \neq i}}^n (\alpha_i - \alpha_j))^{-1}$.

\section{Secure Distributed Matrix Multiplication}
\label{sec:SDMM}

The aim of secure distributed matrix multiplication is to distribute the computation of a matrix product to workers such that the workers do not gain any information about the secret matrices.
The computation should be made redundant such that the responses from some of the workers may be ignored, which will mitigate the so-called straggler problem.

\subsection{System Model}
\label{ssec:system_model}

We wish to compute the product $AB$ for matrices $A \in \F_q^{t \times s}$ and $B \in \F_q^{s \times r}$.
We begin by partitioning these matrices to $P$ equal pieces such that
\begin{equation*}
    A = \begin{pmatrix}
        A_1 & \cdots & A_P
    \end{pmatrix}, \quad B = \begin{pmatrix}
        B_1 \\ \vdots \\ B_P
    \end{pmatrix}.
\end{equation*}
Then, the product can be expressed as
\begin{equation*}
    AB = \sum_{j=1}^P A_jB_j.
\end{equation*}
This way of partitioning the matrices is known as the \emph{inner product partition}, since the product is expressed as the inner product of the block vectors.
Other ways to partition the matrices have also been considered in the literature, but we will not focus on these in this paper.

Each of the $N$ workers, indexed by $i \in [N]$, is sent two matrices $\tilde{A}_i$ and $\tilde{B}_i$.
These encoded matrices depend on the blocks of $A$ and $B$, as well as some randomness.
The workers compute $\tilde{A}_i \tilde{B}_i$ and return this to the user who decodes the product $AB$ from the responses.
The \emph{recovery threshold} of the scheme is the minimal number $R$ such that the product can be decoded from \emph{any} $R$ of the responses.
In some cases, it may be possible to decode the product from some fewer number of responses.

Let $\mathcal{X} \subseteq [N]$, $\lvert \mathcal{X} \rvert \leq X$, be a set of colluding workers and $\tilde{A}_\mathcal{X} = \{\tilde{A}_i : i \in \mathcal{X}\}$, $\tilde{B}_\mathcal{X} = \{\tilde{B}_i : i \in \mathcal{X}\}$ be the collections of the shares of the matrices $A$ and $B$ held by these workers.
The encoded pieces should have the property that
\begin{equation*}
    I(A, B; \tilde{A}_\mathcal{X}, \tilde{B}_\mathcal{X}) = 0,
\end{equation*}
where the mutual information is computed over the randomness in the encoding process.
The parameter $X$ denotes the security level of the system as the security condition describes that any $X$ workers should gain no information about the matrices $A$ and $B$ from their encoded pieces.
If the above condition holds, then we call the SDMM scheme \emph{$X$\nobreakdash-secure}.

The following examples will show how SDMM schemes have been constructed using polynomial evaluation.

\begin{example}[DFT scheme \cite{mital2022secure}]
\label{ex:DFT}
Consider the functions
\begin{align*}
    f(x) &= \sum_{j=1}^P A_j x^{j - 1} + \sum_{k=1}^X R_k x^{P + k - 1}, \\
    g(x) &= \sum_{j'=1}^P B_{j'} x^{-j' + 1} + \sum_{k'=1}^X S_{k'} x^{-P - X - k' + 1},
\end{align*}
where $R_1, \dots, R_X$ and $S_1, \dots, S_X$ are matrices of suitable size chosen uniformly at random.
Each worker is sent $\tilde{A}_i = f(\alpha_i)$ and $\tilde{B}_i = g(\alpha_i)$.
The workers compute $h(\alpha_i) = f(\alpha_i)g(\alpha_i)$, where
\begin{align*}
    h(x) = f(x)g(x) = \sum_{j=1}^P A_jB_j + (\text{non-constant terms}).
\end{align*}
The terms in the polynomial (in $x$ and $x^{-1}$) have degrees in $-(P + 2X) + 1, \dots, P + X - 1$.
Let $N = P + 2X$ and choose the evaluation points $\alpha_i$ to be distinct $N$th roots of unity.
It is well known that
\begin{equation*}
    \sum_{i=1}^N \alpha_i^\ell = \begin{dcases*}
        N & if $N \mid \ell$ \\
        0 & otherwise
    \end{dcases*}.
\end{equation*}
Therefore,
\begin{equation*}
    \frac{1}{N}\sum_{i=1}^N h(\alpha_i) = \sum_{j=1}^P A_j B_j = AB,
\end{equation*}
since the non-constant terms add up to zero.
The existence of the $N$th roots of unity requires that $N \mid (q - 1)$.
\end{example}

\begin{example}[Secure MatDot \cite{lopez2022secure}]
\label{ex:secure_MatDot}
Let $\beta_1, \dots, \beta_P$ and $\alpha_1, \dots, \alpha_{q - P}$ be distinct points in $\F_q$.
Choose polynomials $f(x), g(x) \in \F_q[x]$ of degree $< P + X$ such that
\begin{align*}
    f(\beta_j) &= A_j, \quad g(\beta_j) = B_j, \quad j \in [P] \\
    f(\alpha_k) &= R_k, \quad g(\alpha_k) = S_k, \quad k \in [X].
\end{align*}
Let $N = q - P$.
Worker $i \in [N]$ receives $\tilde{A}_i = f(\alpha_i)$ and $\tilde{B}_i = g(\alpha_i)$.
They compute $h(\alpha_i) = f(\alpha_i)g(\alpha_i)$, where $h(x) = f(x)g(x)$ and $\deg(h(x)) < 2P + 2X - 1$.
Let $\gamma = (\beta_1, \dots, \beta_P, \alpha_1, \dots, \alpha_{q - P})$ be a vector whose entries are all the elements in $\F_q$.
Then,
\begin{equation*}
    \RS_{2P + 2X - 1}(\gamma)^\perp = \RS_{q - 2P - 2X + 1}(\gamma).
\end{equation*}
Let $H(x) \in \F_q[x]$ be such that $\deg(H(x)) < q - 2P - 2X + 1$, $H(\beta_1) = \dots = H(\beta_P) = \mu \neq 0$, and $H(x)$ has many zeros in $\F_q$.
The evaluation vector of $H(x)$ on the evaluation points $\gamma$ is contained in $\RS_{q - 2P - 2X + 1}(\gamma)$, so
\begin{equation*}
    \sum_{j=1}^P h(\beta_j)H(\beta_j) + \sum_{i=1}^N h(\alpha_i)H(\alpha_i) = 0
\end{equation*}
due to the orthogonality relation.
Thus,
\begin{equation*}
    AB = \sum_{j=1}^P A_j B_j = \sum_{j=1}^P h(\beta_j) = -\frac{1}{\mu} \sum_{i=1}^N h(\alpha_i)H(\alpha_i).
\end{equation*}
If $H(\alpha_i) = 0$ for many of the $\alpha_i$, then only a few of the $h(\alpha_i)$ are needed to compute the above sum.
On the other hand, as $h(x)$ has degree $< 2P + 2X - 1$, any $2P + 2X - 1$ responses are sufficient to interpolate $h(x)$ and compute the sum.
The authors of \cite{lopez2022secure} construct suitable polynomials $H(x)$ under several assumptions of the field size dividing some of the parameters.
\end{example}

\subsection{Linear SDMM}
\label{ssec:linear_SDMM}

A general framework for SDMM schemes utilizing linear codes, called \emph{linear SDMM}, was formulated in \cite{makkonen2022general}.
Most of the SDMM schemes presented in the literature can be expressed in terms of this framework, including \cref{ex:DFT,ex:secure_MatDot}.
The encodings are computed using generator matrices $F$ and $G$ of $[N, P + X]$ linear codes $\mathcal{C}_A$ and $\mathcal{C}_B$, respectively.
In particular,
\begin{align*}
    (\tilde{A}_1, \dots, \tilde{A}_N) &= (A_1, \dots, A_P, R_1, \dots, R_X)F, \\
    (\tilde{B}_1, \dots, \tilde{B}_N) &= (B_1, \dots, B_P, S_1, \dots, S_X)G,
\end{align*}
where $R_1, \dots, R_X$ and $S_1, \dots, S_X$ are random matrices of suitable size chosen uniformly at random.
The responses from the workers are
\begin{equation*}
    (\tilde{A}_1 \tilde{B}_1, \dots, \tilde{A}_N \tilde{B}_N) \in \mathcal{C}_A \star \mathcal{C}_B.
\end{equation*}
The subcodes of $\mathcal{C}_A$ and $\mathcal{C}_B$ that correspond to the random parts are denoted as $\mathcal{C}_A^\sec$ and $\mathcal{C}_B^\sec$.
These codes are generated by the lowest $X$ rows of the generator matrices $F$ and $G$, respectively.
The following well-known proposition is used to show the security of most SDMM schemes in the literature \cite[Theorem 1]{makkonen2022general}.

\begin{proposition}[Security of linear SDMM]
\label{prop:MDS_security}
A linear SDMM scheme is $X$\nobreakdash-secure if $\mathcal{C}_A^\sec$ and $\mathcal{C}_B^\sec$ are $[N, X]$ MDS codes.
\end{proposition}

By \cite[Theorem 3]{makkonen2022general}, $N \geq P + 2X$ for any linear SDMM scheme with $\mathcal{C}_A^\sec$ and $\mathcal{C}_B^\sec$ MDS codes.
If $X \geq q + 1$, then $N \leq X + 1$, which is a contradiction, so $X \leq q$.
Assuming that the MDS conjecture holds and $X \geq 2$, then $q \geq N - 1$ (or $q \geq N - 2$), where $q$ is the field size and $N$ is the number of workers.
According to this, it would not be possible to construct $X$\nobreakdash-secure SDMM schemes over field sizes significantly smaller than the number of workers\footnote{
The scheme in \cite{machado2023hera} is proposed to work over small field sizes (even sublinear in~$N$) if the MDS condition in their generator matrices is fulfilled.
However, this seems to not be possible in generality, assuming the MDS conjecture holds.
}.
For $X = 1$, it may still be possible to reduce the field size due to the fact that the $[N, 1]$ repetition code is MDS over any field.

\section{Construction}
\label{sec:construction}

Denote the length $P$ row vectors whose entries are the matrix partitions by $a = (A_1, \dots, A_P)$ and $b = (B_1, \dots, B_P)$.
We have that $ab^T = AB$ due to the inner product partitioning.
If $M$ is an invertible $P \times P$ matrix and we are able to compute $aMb^T$ for all $a$ and $b$, then we may simply compute $(aM^{-1})Mb^T = ab^T = AB$.
This corresponds to first doing a linear transformation on the partitions of matrix $A$.
This allows us to consider the simpler problem of computing $aMb^T$ for some fixed, but arbitrary, invertible $P \times P$ matrix~$M$.

Let $q \geq N$ and $\alpha \in \F_q^N$ be a vector with distinct entries.
Consider the encoding polynomials
\begin{align*}
    f(x) &= \sum_{k=1}^{X} R_k x^{k - 1} + \sum_{j=1}^{P} A_j x^{X + j - 1}, \\
    g(x) &= \sum_{k'=1}^{X} S_{k'} x^{k' - 1} + \sum_{j'=1}^{P} B_{j'} x^{X + j' - 1},
\end{align*}
where $R_1, \dots, R_X$ and $S_1, \dots, S_X$ are matrices of the same size as the partitions $A_j$ and $B_{j'}$ chosen uniformly at random.
We define $\tilde{A}_i = f(\alpha_i), \tilde{B}_i = g(\alpha_i)$ for $i \in [N]$.
By definition of Reed--Solomon codes, we have that
\begin{equation*}
    \tilde{A} \in \mathcal{C}_A = \RS_{P + X}(\alpha), \quad \tilde{B} \in \mathcal{C}_B = \RS_{P + X}(\alpha).
\end{equation*}
Furthermore, the security codes are
\begin{equation*}
    \mathcal{C}_A^\sec = \mathcal{C}_B^\sec = \RS_X(\alpha).
\end{equation*}
As these codes are MDS, we know that this scheme is $X$\nobreakdash-secure according to \cref{prop:MDS_security}.

The workers compute $\tilde{A}_i \tilde{B}_i$, which means that we receive evaluations of the polynomial $h(x) = f(x)g(x)$.
These response vectors are contained in the star product code
\begin{equation*}
    \mathcal{C}_A \star \mathcal{C}_B = \RS_{\min\{2P + 2X - 1, N\}}(\alpha).
\end{equation*}
The interference terms of $h(x)$ (those including random parts) are contained in the terms of degree $0, \dots, P + 2X - 2$.
Therefore,
\begin{equation*}
    h(x) = (\text{terms of degree $< P + 2X - 1$}) + \sum_{j=1}^{P} \sum_{j'=1}^{P} A_j B_{j'} x^{2X + j + j' - 2}
\end{equation*}
Let us choose a decoding vector $\lambda \in \RS_{P + 2X - 1}(\alpha)^\perp$, \emph{i.e.},
\begin{equation*}
    \sum_{i=1}^{N} \lambda_i \alpha_i^\ell = 0
\end{equation*}
for $\ell = 0, \dots, P + 2X - 2$.
Then,
\begin{align*}
    \sum_{i=1}^{N} \lambda_i h(\alpha_i) &= \sum_{j=1}^P \sum_{j'=1}^{P} A_j B_{j'} \sum_{i=1}^{N} \lambda_i \alpha_i^{2X + j + j' - 2},
\end{align*}
as all terms of degree $< P + 2X - 1$ add up to zero according to the definition of $\lambda$.
We call this interference cancellation.

We can further write the above sum as $aMb^T$, where the $P \times P$ matrix $M$ is defined by
\begin{equation*}
    M_{j,j'} = \sum_{i=1}^{N} \lambda_i \alpha_i^{2X + j + j' - 2}.
\end{equation*}

\begin{lemma}
If $\lambda \in \RS_{P + 2X - 1}(\alpha)^\perp \setminus \RS_{P + 2X}(\alpha)^\perp$, then the matrix $M$ is invertible.
\end{lemma}

\begin{IEEEproof}
Notice that $M_{j, j'}$ only depends on $j + j'$, so let $m_{j + j'} = M_{j, j'}$.
As $\lambda \in \RS_{P + 2X - 1}(\alpha)^\perp$, we have that $m_\ell = 0$ for $\ell < P + 1$.
Therefore, the matrix $M$ has the following form
\begin{align*}
    M = \begin{pmatrix}
        0 & & m_{P + 1} \\
        & \iddots & \\
        m_{P + 1} & & \star
    \end{pmatrix}.
\end{align*}
Thus, $M$ is invertible if and only if $m_{P + 1} \neq 0$.
If $m_{P + 1} = 0$, then $\sum_{i=1}^{N} \lambda_i \alpha_i^\ell = 0$ for $\ell = 0, \dots, P + 2X - 1$, which implies that $\lambda \in \RS_{P + 2X}(\alpha)^\perp$.
This is a contradiction to the definition of $\lambda$, which means that $m_{P + 1} \neq 0$.
Therefore, $M$ is invertible.
\end{IEEEproof}

Using the above lemma, we can compute
\begin{equation*}
    \sum_{i=1}^{N} \lambda_i h(\alpha_i) = aMb^T,
\end{equation*}
where $M$ is invertible.
The only requirement on the field size is that $q \geq N$ for the existence of the Reed--Solomon code of length $N$.
In the following, we present two constructions based on this idea.

\begin{theorem}[Construction without redundancy]
\label{thm:construction_without_redundancy}
Let $N = P + 2X$ and $q \geq N$.
Then there exists an $X$\nobreakdash-secure SDMM scheme over $\F_q$ using $N$ workers.
\end{theorem}

\begin{IEEEproof}
By properties of Reed--Solomon codes,
\begin{align*}
    \RS_{P + 2X - 1}(\alpha)^\perp &= \spn\{\omega\} \\
    \RS_{P + 2X}(\alpha)^\perp &= \{0\},
\end{align*}
for some $\omega \in (\F_q^*)^N$.
Therefore, we need to choose $\lambda \in \spn\{\omega\} \setminus \{0\}$.
\end{IEEEproof}

We may also add redundancy to the construction such that some of the workers may be ignored.
Let $S$ denote the number of straggling workers we wish to add resiliency to.

\begin{theorem}[Construction with redundancy]
\label{thm:construction_with_redundancy}
Let $N = 2P + 2X + S - 1$ and $q \geq N$.
Then there exists an $X$\nobreakdash-secure SDMM scheme over $\F_q$ using $N$ workers such that the product can be decoded from any $2P + 2X - 1$ workers or from some specified $P + 2X$ workers.
\end{theorem}

\begin{IEEEproof}
By properties of Reed--Solomon codes,
\begin{align*}
    \RS_{P + 2X - 1}(\alpha)^\perp &= \GRS_{P + S}(\alpha, \omega) \\
    \RS_{P + 2X}(\alpha)^\perp &= \GRS_{P + S - 1}(\alpha, \omega),
\end{align*}
for some $\omega \in (\F_q^*)^N$.
The minimum distance of the first code is $P + 2X$, while the minimum distance of the second code is $P + 2X + 1$.
Let $\lambda$ be a nonzero codeword of minimum weight in $\RS_{P + 2X - 1}(\alpha)^\perp$.
Then it is clear that $\lambda \notin \RS_{P + 2X}(\alpha)^\perp$.
As $\wt(\lambda) = P + 2X$, it is enough to receive some specific $P + 2X$ responses, since
\begin{equation*}
    aMb^T = \sum_{i=1}^P \lambda_i h(\alpha_i) = \:\: \sum_{\mathclap{i \in \supp(\lambda)}} \:\: \lambda_i h(\alpha_i).
\end{equation*}

As the responses are contained in the code $\mathcal{C}_A \star \mathcal{C}_B = \RS_{2P + 2X - 1}(\alpha)$, it is enough to receive \emph{any} $2P + 2X - 1 = N - S$ responses to decode all $h(\alpha_i)$ and then compute
\begin{equation*}
    aMb^T = \sum_{i=1}^{N} \lambda_i h(\alpha_i). \IEEEQEDhereeqn
\end{equation*}
\end{IEEEproof}

As the dual of a Reed--Solomon code can be explicitly computed, it is easy to find explicit constructions for the above theorems.
The encoding process and the decoding process are simply linear combinations of the matrix blocks.

\subsection{Construction Over the Binary Field}
\label{ssec:construction_over_the_binary_field}

As the MDS conjecture does not state anything about MDS codes of dimension one, such as the repetition code, we may construct the following SDMM scheme over a small field.
Let $P$ be even, $X = 1$, $N = P + 2$, and work over $\F_2$.
Consider the following shares:
\begin{align*}
    \tilde{A}_j &= R + A_j, & \tilde{B}_j &= S + B_j, \quad j \in [P] \\[-.5ex]
    \tilde{A}_{P + 1} &= R + \sum_{j=1}^P A_j, & \tilde{B}_{P + 1} &= S, \\[-3ex]
    \tilde{A}_{P + 2} &= R, & \tilde{B}_{P + 2} &= S + \sum_{j=1}^P B_j.
\end{align*}
Then, the sum of the responses $\tilde{A}_i \tilde{B}_i$, $i \in [N]$, is
\begin{align*}
    &\quad\sum_{j=1}^P (R + A_j)(S + B_j) + \bigg( \! R + \sum_{j=1}^P A_j \! \bigg) S + R \bigg( \! S + \sum_{j=1}^P B_j \! \bigg) \\
    &= \sum_{j=1}^P A_jB_j + \bigg( \sum_{j=1}^P A_j \bigg) S + R \bigg( \sum_{j=1}^P B_j \bigg) + P \cdot RS \\
    &+ RS + \bigg( \sum_{j=1}^P A_j \bigg) S + RS + R \bigg( \sum_{j=1}^P B_j \bigg) = AB.
\end{align*}
It is also clear that this is secure with $X = 1$, since all $\tilde{A}_i$ and $\tilde{B}_i$ are protected with uniform noise.

\section{Comparison}
\label{sec:comparison}

\begin{table*}[!t]
    \renewcommand{\arraystretch}{1.3}
    \caption{Comparison between different inner product partitioning SDMM schemes}
    \label{tab:comparison}
    \centering
    \scriptsize
    \begin{tabular}{|c|c|c|c|c|c|}
        \hline

        \textbf{Construction} & \textbf{Stragglers} & \textbf{Number of workers} & \textbf{Recovery threshold} & \textbf{Minimal recovery size} & \textbf{Field size conditions} \\ \hhline{|=|=|=|=|=|=|}

        \cref{thm:construction_without_redundancy} & No & $P + 2X$ & $P + 2X$ & $P + 2X$ & $q \geq N$ \\ \hline

        \cref{thm:construction_with_redundancy} & Yes & $2P + 2X + S - 1$ & $2P + 2X - 1$ & $P + 2X$ & $q \geq N$ \\ \hline

        Secure MatDot \cite{lopez2022secure} & Yes & $2P + 2X + S - 1$ & $2P + 2X - 1$ & $\geq P + 2X$ & $q \geq P + N$, divisibility conditions \\ \hline

        HerA \cite{machado2023hera} & No & $P + 2X$ & $P + 2X$ & $P + 2X$ & $q$ is a square, $q^{3/2} \geq 2(P + X)$ \\ \hline

        DFT \cite{mital2022secure} & No & $P + 2X$ & $P + 2X$ & $P + 2X$ & $N \mid (q - 1)$ \\ \hline
    \end{tabular}
\end{table*}

In this section we will compare the constructions given in \cref{thm:construction_without_redundancy,thm:construction_with_redundancy} to those in \cite{machado2023hera,lopez2022secure,mital2022secure}.
In particular, we will compare the number of workers required $N$, the recovery threshold $R$, the minimal number of workers needed for decoding, and the field size requirements.
These are listed in \cref{tab:comparison}.

The authors of \cite{machado2023hera} do not show the existence of their construction for general parameters $P$ and $X$.
As commented earlier, under the MDS security condition (\cref{prop:MDS_security}) and the MDS conjecture, it is not possible to reduce the field size to below $N - 2$ even by using algebraic geometry codes, when $X > 1$.
On the other hand, the secure MatDot scheme in \cite{lopez2022secure} requires the existence of a certain subgroup to make their construction work, which requires some divisibility constraints on the field size $q$ and the other parameters.
Finally, the DFT scheme assumes $N \mid (q - 1)$, which restricts the possible field.

The constructions given in \cref{thm:construction_without_redundancy,thm:construction_with_redundancy} provide the most flexibility in the field size and almost reach the bound $q \geq N - 1$ (or $q \geq N - 2$) given by the MDS conjecture.
Furthermore, \cref{thm:construction_with_redundancy} achieves a balance between having redundancy against any $S$ straggling workers, as well as having a set of few workers whose responses are sufficient for recovering the product.

This paper has focused on the number of workers, the recovery threshold, and the minimal recovery size, but for some applications the total download cost is more important.
The authors of \cite{machado2021field} devise an SDMM scheme using the inner product partitioning that is able to reach lower download cost by downloading symbols in a subfield from a larger number of total workers.
This construction requires extremely large field extensions as seen in \cite[Section VI]{machado2021field}.

\section{Conclusions}

In this paper, we construct SDMM schemes with minimal field size restrictions compared to the previous literature.
This construction can be seen as a generalization of the DFT scheme presented in \cite{mital2022secure} as we are able to efficiently cancel all interference terms in a single linear combination.
We discuss the implications of the MDS conjecture on the field size requirements of $X$\nobreakdash-secure SDMM schemes and find that achieving significantly smaller field sizes is not possible.

So-called extended Reed--Solomon codes achieve the bound $n = q + 1$ given by the MDS conjecture.
It would be interesting to extend the construction given in this paper to work with extended Reed--Solomon codes and achieve the lowest possible field sizes for $X$\nobreakdash-secure SDMM schemes.

\section*{Acknowledgment}
\addcontentsline{toc}{section}{Acknowledgments}

This work has been supported by the Research Council of Finland under Grant No.\ 336005 (PI C.~Hollanti) and by the Vilho, Yrjö and Kalle Väisälä Foundation of the Finnish Academy of Science and Letters.

\bibliographystyle{IEEEtran}
\bibliography{bib}

\end{document}